\documentclass[conference]{IEEEtran}
\IEEEoverridecommandlockouts

\ifCLASSOPTIONcompsoc
\usepackage[caption=false, font=normalsize, labelfont=sf, textfont=sf]{subfig}
\else
\usepackage[caption=false, font=normalsize]{subfig}
\fi
\usepackage{lipsum}%
\usepackage[dvipsnames]{xcolor}
\usepackage{algorithm,algorithmic}
\usepackage{balance}
\usepackage{multicol}   
\usepackage{cite}
\usepackage{gensymb}
\usepackage{multirow}
\usepackage{graphics}
\usepackage{epsfig}
\usepackage{graphicx}
\usepackage{epstopdf}
\usepackage{textcomp}
\usepackage{amsmath}
\usepackage{mathtools}
\usepackage{filecontents}
\usepackage{lipsum,color}
\usepackage{amssymb}
\usepackage{float}
\usepackage{colortbl} 
\usepackage{times} 
\usepackage{amsthm}  

\usepackage{amsfonts}

\usepackage{bm}
\SetSymbolFont{largesymbols}{bold}{OMX}{txex}{b}{n}

\theoremstyle{break}

\begin{document}

\title{UAV Trajectory Optimization for Directional THz Links Using DRL
}

\author{\IEEEauthorblockN{ Mohammad~Taghi~Dabiri,~and~Mazen~Hasna}
	\IEEEauthorblockA{\textit{Department of Electrical Engineering,} 
		\textit{Qatar University}, Doha, Qatar, \\
		E-mails: (m.dabiri@qu.edu.qa; hasna@qu.edu.qaa).}
	\thanks{This publication was made possible by grant number NPRP13S-0130- 200200 from the Qatar National Research Fund, QNRF. The statements made herein are solely the responsibility of the authors.}
}

\maketitle

\begin{abstract}
	As an alternative solution for quick disaster recovery of backhaul/fronthaul links, in this paper, a dynamic unmanned aerial vehicles (UAV)-assisted heterogeneous (HetNet) network equipped with directional terahertz (THz) antennas is studied to solve the problem of transferring traffic of distributed small cells. To this end, we first characterize a detailed three-dimensional modeling of the dynamic UAV-assisted HetNet, and then, we formulate the problem for UAV trajectory to minimize the maximum outage probability of directional THz links. Then, using deep reinforcement learning (DRL) method, we propose an efficient algorithm to learn the optimal trajectory. Finally, using simulations, we investigate the performance of the proposed DRL-based trajectory method.
\end{abstract}

\IEEEpeerreviewmaketitle

\section{Introduction}
Unmanned aerial vehicle (UAV)-assisted backhaul/fronthaul links are proposed as an alternative easy to deploy solution for quick network recovery after disasters when the network infrastructure (mainly is based on fragile optical fiber links) is out of access.
Microwave backhaul/fronthaul links can cover a wide area but suffer from low data rates. The high frequency millimeter wave (mmWave) and terahertz (THz) links meet the capacity requirements of next generation communication networks. 
However, in a dynamic network, the design of the UAV-based network with THz links is complicated, and the UAVs should adjust their positions in the three-dimensional (3D) space in relation to the distributed dense small cell base stations (SBSs) in such a way that the interference between the THz links is reduced.
UAV trajectory with the help of reinforcement learning (RL) algorithms can provide a reliable service for distributed users, which is the subject of several recent works \cite{maleki2022multi, zeng2021simultaneous, fonseca2023adaptive, bellone2022deep, gopi2021reinforcement, hoseini2020trajectory, chen2022trajectory, chen2021joint, chen2021multi, zhang2022cooperative}.
However, the results of these works are not suitable for a UAV-based network that uses directional THz links. Due to the small beamwidth of the directional THz links, the small fluctuations of the UAV, even in the order of one degree, can affect the performance of the system, and therefore the THz beam width cannot be chosen too small, which leads to interference between THz links. In this case, during the trajectory, the UAV must simultaneously adjust its antenna pattern to control the interference between randomly distributed nodes which is the subject of this work. 

\begin{figure}
	\begin{center}
		\includegraphics[width=3.4 in]{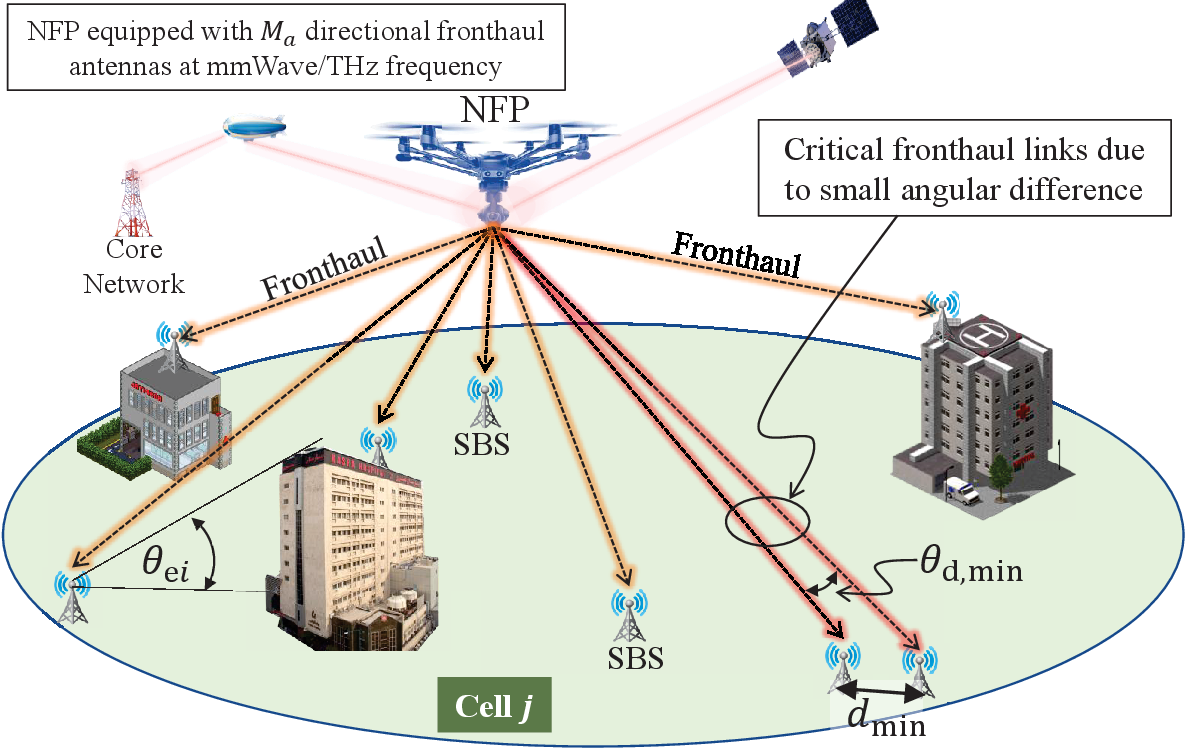}
		\caption{An illustration of a UAV-assisted HetNet as an alternative solution for fronthaul links which uses directional THz antenna to transfer traffic from the distributed SBSs to the core network.}
		\label{se1}
	\end{center}
\end{figure}

In this study, we consider a dynamic UAV-assisted heterogeneous network (HetNet) as shown in Fig. \ref{se1} that is offered as an easy to deploy solution of fronthaul links to solve the problem related to transferring traffic of the distributed SBSs to the core network.  
First, we characterize a detailed 3D modeling of the dynamic UAV-assisted HetNet, by taking into account the random distribution of SBSs, spatial angles between THz links, real antenna pattern, and UAV's vibrations in the 3D space. Using this characterization, we formulate the problem for the UAV trajectory in a dynamic network.
Then a deep RL framework is proposed to solve the trajectory problem of the dynamic network with the objective of minimizing the maximum outage probability of fronthaul links.
Finally, by providing simulations, we will investigate the performance of the proposed algorithm in different scenarios.





\section{The System Model}

As shown in Fig. \ref{se1}, we consider a UAV-assisted HetNet that as an alternative for fronthaul links, the UAV-based THz links are used to transfer the SBSs traffic \cite{9998554,9904477}.
We assume UAV (in this work, UAV and NFP both refer to the same concept) is equipped with $M_u$ directional antennas denoted by $A_i$ where $i\in\{1,...,M_u\}$.
The direction of each $A_i$ is set towards a SBS assigned to it. 
%
SBSs are randomly distributed in a two-dimensional ground space and they can change randomly over time.
Let's respectively denote the number and position of the SBSs by $M_s$ and $S_i$ where $i\in\{1,...,M_s\}$. $M_s\in\{M_\text{s,min},...,M_\text{s,max}\}$ is a uniform discrete random variable (RV) where $M_\text{s,max}=M_u$.
%
%
The position of each $S_i$ is characterized as $[x_i,y_i,0]$ in a Cartesian coordinate system where $[x,y,z]\in\mathbb{R}^{1\times3}$. Also, $F_i$ stands the fronthaul link between $A_i$ and $S_i$.

We consider a dynamic general network in which the number and distribution of SBSs changes randomly over time.
We assume that the the number and distribution of SBSs changes with time $T\in[T_\text{min},T_\text{max}]$ which is a uniform RV. 
The changes in the network after time $T$ are such that $m_d$ of the SBSs are randomly disconnected and the other $m_c$ SBSs are connected to the UAV with new random positions on the $x-y$ plane. 
The parameters $m_d\in\{0,...,m_{d,\text{max}}\}$ and $m_c\in\{0,...,m_{c,\text{max}}\}$ are RVs where $m_{c,\text{max}} = M_s-m_d$.

With any change in network topology, the UAV must continuously modify its position.
We assume that the UAV flies with the maximum speed constraint $v_\text{max}$ in (m/s) and the maximum acceleration constraint $v'_\text{max}$ in (m/s$^2$).	
Let $B_u(t)=[x_u(t),y_u(t),z_u(t)]$ represent the instantaneous position of UAV at time $t$. For a short time $\Delta t$, the position of UAV is updated as
\begin{align} \label{aa1}
\left\{ \!\!\!\!\! \! \!
\begin{array}{rl}
	& x_u(t+\Delta t) = x_u(t) + v_x(t)\Delta t + \frac{1}{2} v'_x(t) \Delta t^2, \\
	& y_u(t+\Delta t) = y_u(t) + v_y(t)\Delta t + \frac{1}{2} v'_y(t) \Delta t^2, \\
	& z_u(t+\Delta t) = z_u(t) + v_z(t)\Delta t + \frac{1}{2} v'_z(t) \Delta t^2, 
\end{array} \right.
\end{align}
where $v(t)=[v_x(t)],v_y(t),v_z(t)]$ and $v'(t)=[v'_x(t)],v'_y(t),v'_z(t)]$	are instantaneous speed and acceleration of UAV, respectively. 
For notation simplicity, we will remove the notation $t$ of $B_u(t)=[x_u(t),y_u(t),z_u(t)]$ in the following, except where necessary.

We consider a uniform square array antennas for both UAV and SBSs.
Let $N_{ui}\times N_{ui}$ represent antenna elements of $A_i$ for $i\in\{1,...,M_u\}$ with the same spacing $d_a$ between elements. Similarly, $N_{si}\times N_{si}$ is antenna elements of $S_i$ for $i\in\{1,...,M_s\}$.
The array radiation gain is mainly formulated in the direction of $\theta$ and $\phi$, where $\theta$ and $\phi$ are clearly defined in \cite[Fig. 6.28]{balanis2016antenna}.
By taking into account the effect of all elements, the array radiation gain will be:
\begin{align}
\label{p_1}
G_{qi}(N_{qi},\theta,\phi)  = G_0(N_{qi}) 
G_{ai}(N_{qi},\theta,\phi),
\end{align}
where $G_{ai}$ is an array factor and $G_0$ is defined in \eqref{cv}. Also, the subscript $q=s$ determines the antenna of $S_i$ and the subscript $q=u$ determines the antenna $A_i$.  
If the amplitude excitation of the entire array is uniform, then the array factor $G_{ai}(N_{qi},\theta,\phi)$ for a square array of $N_{qi}\times N_{qi}$ elements can be obtained as \cite[eqs. (6-89) and (6-91)]{balanis2016antenna}:
\begin{align}
\label{f_1}
G_{ai}(N_{qi},\theta, \phi) &= 
\left( \frac{\sin\left(\frac{N_{qi} (k d_a \sin(\theta)\cos(\phi)+\mathbb{V}_{x})}{2}\right)} 
{N_{qi}\sin\left(\frac{k d_a \sin(\theta)\cos(\phi)+\mathbb{V}_{x}}{2}\right)}
\right. \nonumber \\
&\times \left. \frac{\sin\left(\frac{N_{qi} (k d_a \sin(\theta)\sin(\phi)+\mathbb{V}_{y})}{2}\right)} 
{N_{qi}\sin\left(\frac{k d_a \sin(\theta)\sin(\phi)+\mathbb{V}_{y}}{2}\right)}\right)^2,
\end{align}
where $d_a=\frac{\lambda}{2}$ and $\mathbb{V}_{w}$ are the spacing and progressive phase shift between the elements, respectively. $k=\frac{2\pi}{\lambda}$ is the wave number, $\lambda=\frac{c}{f_c}$ is the wavelength, $f_c$ is the carrier frequency, and $c$ is the speed of light.
Also, in order to guarantee that the total radiated power of antennas with different $N_{qi}$ are the same, the coefficient $G_0$ is defined as
\begin{align}
\label{cv}
G_0(N_{qi})=\frac{4\pi}{\int_0^{\pi}\int_0^{2\pi} G_{ai}(N_{qi},\theta,\phi) \sin(\theta) d\theta d\phi}.
\end{align}
Based on \eqref{f_1}, the maximum value of the antenna gain is equal to $G_0(N_{qi})$, which is obtained when $\theta=0$.
Let $\Theta=[\Theta_x,\Theta_y]$ denote the UAV's orientation fluctuations.
Based on the central limit theorem, the UAV's orientation fluctuations are considered to be Gaussian distributed \cite{dabiri2018channel,dabiri2020analytical}. 
Therefore, we have 
$\Theta_{x}\sim \mathcal{N}(0,\sigma^2_{\theta})$, and $\Theta_{y}\sim \mathcal{N}(0,\sigma^2_{\theta})$.
For the ground SBSs, the estimation error of the exact position of the flying UAV and the insufficient speed to track UAV will cause an angular error \cite{dabiri2022general,dabiri2022pointing}. Unlike the flying UAV, we assume that the ground SBSs do not face weight and power consumption limitations and, hence, they can align their antennas with the considered UAV with a negligible angle error. Therefore, in a real scenario, received power can be obtained as follows:
\begin{align}
\label{pp4}
& P_{r_i} 2 P_{t_i} |h_{L_i}|^2 G_0(N_{si}) G_{uj}(N_{uj},\Theta,\Phi)  \nonumber \\
&+|h_{L_i}|^2 \sum_{j=1,j\neq i}^{M_s} 
P_{t_j}  
G_0(N_{si}) 
G_{uj}(N_{uj},\theta_{ij},\phi_{ij}),
\end{align}
where $\theta_{ij}=[\theta_{x_{ij}}+\Theta_x,\theta_{y_{ij}}+\Theta_y]$,
$P_{t_i}$ is the transmit power of $A_i$, $L_i$ is the linklength $F_i$, 
$h_{L_i}=h_{Lf}(L_i)h_{Lm}(L_i)$ is the channel path loss, 
$h_{Lf}(L_i)=\left(\frac{\lambda}{4\pi L_{i}}\right)^2$ is the free-space path loss, $h_{Lm}(L_i)= e^{-\frac{\mathcal{K}(f)}{2}L_i}$ represents the molecular absorption loss, and $\mathcal{K}(f)$ is the frequency dependent absorption coefficient. Also, the parameter $\phi_{ij}$ is the roll angle of pattern $A_j$ compared to $S_i$, and the parameter $\theta'_{ij} = [\theta_{x_{ij}},\theta_{y_{ij}}]$ is the spatial angle between $F_i$ and $F_j$ links where
\begin{align}
\label{s2}
\left\{ \!\!\!\!\! \! \!
\begin{array}{rl}
	&\theta_{x_{ij}} = \text{cos}^{-1}\left( \frac{(x_u-x_i)^2 + (x_u-x_j)^2+2z_u^2 - d_{x_{ij}}^2}
	{2\sqrt{[(x_u-x_i)^2 +z_u^2] [(x_u-x_j)^2+z_u^2]}}   \right), \\
	& \theta_{y_{ij}} = \text{cos}^{-1}\left( \frac{(y_u-y_i)^2 + (y_u-y_j)^2+2z_u^2 - d_{y_{ij}}^2}
	{2\sqrt{[(y_u-y_i)^2 +z_u^2] [(y_u-y_j)^2+z_u^2]}}   \right), 
\end{array} \right.
\end{align}
and $d_{x_{ij}}=|x_i-x_j|$ and $d_{y_{ij}}=|y_i-y_j|$.

%
Therefore, the probability of LoS is an important factor and can be described as a function of the elevation angle and environment as follows \cite{al2014modeling,al2014optimal}:
\begin{align}
\label{op11}
P_\textrm{LoS}(\theta_{ei}) = \frac{1}
{1+\alpha \exp\left(-b(\frac{180}{\pi}\theta_\text{ei}-\alpha)\right)}	
\end{align}
where $\alpha$ and $b$ are constants whose values depend on the propagation environment, e.g., rural, urban, or dense urban, and $\theta_\textrm{ei}$ is the elevation angle of $S_i$ compared to the instantaneous position of UAV and can be formulated as
\begin{align}
\label{pp5}
\theta_{ei} = \tan^{-1}\left( \frac{z_u}{\sqrt{(x_i-x_u)^2+(y_i-y_u)^2}} \right).
\end{align}
Finally, the SINR is modeled as
\begin{align}
\label{si1}
\gamma_i = \frac{P_{t_i} \alpha_{L_i} |h_{L_i}|^2 G_0(N_{si}) G_{uj}(N_{uj},\Theta,\Phi)  }
{ \sum_{j=1,j\neq i}^{M_s} 
	P_{t_j}  \alpha_{L_i} |h_{L_i}|^2 G_0(N_{si}) 
	G_{uj}(N_{uj},\theta_{ij},\phi_{ij}) 
	+ \sigma^2_{N}  }, 
\end{align}
where $\sigma^2_{N}$ is the thermal noise power, and coefficient $\alpha_{L_i}$ determines $S_i$ is in the LoS or NLoS of UAV.

\section{Problem Formulation}
Distribution of active fronthaul links connected to the UAV changes with time $T$, which is a random parameter.
Let $\mathcal{B}_u(\mathcal{T})=\{B_u(t_0),B_u(t_0+\Delta t),...,B_u(t_0+J_a\Delta t)\}$ represents a set of UAV movements in the time period $\mathcal{T}=[t_0,t_0+J_a\Delta t]$,
where $B_u(t+j\Delta t)=[x_u(t+j\Delta t),y_u(t+j\Delta t),z_u(t+j\Delta t)]$ for $j\in\{0,1,...,J_a\}$, and $J_a$ is the number of UAV's actions.
The trajectory time that the UAV takes to satisfy the requested QoS of all fronthaul links is denoted as $\mathbb{T}_{ep}$ and is defined as:
\begin{align}
\label{g1}
&\mathbb{T}_{ep} = \sum_{j=1}^{J_a} 
\frac{\Delta B_u(t_0+j\Delta t)}{v(t_0+j\Delta t)},
\end{align}
where
\begin{align}
&\Delta B_u(t_0+j\Delta t) = \nonumber  \\
&\sqrt{\Delta^2 x_u(t_0+j\Delta t)  +  \Delta^2 y_u(t_0+j\Delta t)  +  \Delta^2 z_u(t_0+j\Delta t)}, \nonumber
\end{align}
and
$\Delta x_u(t_0+j\Delta t) = x_u(t+(j+1)\Delta t)-x_u(t+j\Delta t)$,
$\Delta y_u(t_0+j\Delta t) = y_u(t+(j+1)\Delta t)-y_u(t+j\Delta t)$, and
$\Delta z_u(t_0+j\Delta t) = z_u(t+(j+1)\Delta t)-z_u(t+j\Delta t)$.
%
Notice, the trajectory time is more important because we have a dynamic network where on average, the topology of the network changes every $\bar{T}=\frac{T_\text{max}+T_\text{min}}{2}$ second.
Regarding the trajectory problem, the UAV seeks to find the optimal trajectory in a minimum time $\mathbb{T}_{ep}$ under the subject that $\mathbb{T}_{ep}\leq\bar{T}$. The optimization problem can be formulated as
\begin{align}
\label{g3}
& \underset{ \mathcal{B}_u(\mathcal{T}),   \mathcal{N}'_u(\mathcal{T}),
	\mathcal{P}'_t(\mathcal{T})}
{\min} ~~~~
\mathbb{T}_{ep},  \\
& ~~~~~~~~~~~~~~~~~~\text{s.t.}~~~~ \mathbb{T}_{ep}\leq\bar{T} . \nonumber
\end{align}
%
To achieve fair performance among all fronthaul links, we want to minimize the maximum OP over all SBSs
where op is obtained as
\begin{align}
\label{po1}
\mathbb{P}_{\text{out},i}= \text{Prob}\left[\gamma_{i}<\gamma_\text{th}\right].
\end{align}
Therefore, our optimization problem for trajectory is formulated as follows:
\begin{subequations}\label{optim1}
\begin{align}
	\label{opt1}
	& \underset{ \mathcal{B}_u(\mathcal{T}),   \mathcal{N}'_u(\mathcal{T}),
		\mathcal{P}'_t(\mathcal{T})}
	{\min}
	& & \mathbb{T}_{ep}, \\
	\label{opt2}
	& \underset{ \mathcal{B}_u(\mathcal{T}),   \mathcal{N}'_u(\mathcal{T}),
		\mathcal{P}'_t(\mathcal{T})}
	{\min}
	& & \max[P_{\text{out},1},...,P_{\text{out},M_s}] \\
	\label{opt4}
	& ~~~~~~~~~~~~~~\text{s.t.}
	& & P_{\text{out},i}<P_\text{out,th},~~i\in\{1,...,M_s\}, \\
	\label{opt7}
	&&& h_\text{min}\leq z_u(t) \leq h_\text{max}.
\end{align}
\end{subequations}

\section{Analysis and Algorithms}
The optimization problem \eqref{optim1} for UAV trajectory is NP-hard because it is a nonconvex and nonlinear optimization problem \cite{nauss2003solving}.
Therefore, we are not able to solve the optimization problems \eqref{optim1} by classical programming methods and we move to use RL-based methods.

The state space of our optimization problem is a continuous 3D space for the position of the UAV $s_t = B_u(t)=[x_u(t),y_u(t),z_u(t)]$ where $h_\text{min}\leq z_u(t) \leq h_\text{max}$, and the action is also a continuous 3D variable $a_t = [a_{xt}, a_{yt}, a_{zt}]$ where 
\begin{align}
\label{stat_2}
s_{t+1} = \left\{ \!\!\!\!\! \! \!
\begin{array}{rl}
	& s_t + a_t,~~~\text{if}~~~ h_\text{min}<z_u(t)+a_{zt}<h_\text{max},\\
	& s_t,~~~~~~~~~~~~~~\text{otherwise}.
\end{array} \right.
\end{align}
Because the action space is continuous, gradient-based learning algorithms allows us to just follow the gradient to find the best parameters.
Thereby, for the considered UAV-based system, deep deterministic policy gradient (DDPG) algorithm or its variants are fit to find the optimal policy for the agent.
%
While DDPG can achieve great performance sometimes, it is frequently unstable with respect to hyperparameters because there is a risk of overestimating Q-values in the critic (value) network \cite{Twin_Delayed_DDPG}.
Twin Delayed DDPG (TD3) is an efficient policy gradient algorithm that addresses this issue by introducing several critical tricks \cite{Twin_Delayed_DDPG}.
In this paper, TD3 is used to find an optimal policy for the continuous actions of UAV. 
TD3 algorithm consists of two critic deep neural networks (DNNs) $Q(s_t,a_t,\psi_i)$ for $i\in\{1,2\}$, two target DNNs $Q(s_t,a_t,\psi'_i)$ related to the $Q(s_t,a_t,\psi_i)$, one actor DNN $\pi(s_t,\phi)$, and one target DNN $\pi(s_t,\phi'_i)$ related to the $\pi(s_t,\phi_i)$.
At every time training step, TD3 updates the parameters of each critic by minimizing the cost function for training the critic DNN. 
Moreover, every $d_1$ steps, we update the parameters of actor $\phi$ by minimizing the following cost function:
\begin{align}
\label{aa3}
J_\phi   = \sum_s d_\pi(s)  Q(s,\pi(s_t,\phi)+\epsilon;\psi_1),
\end{align}
where $a_t = \pi(s_t,\phi) + \epsilon$ is the final deterministic and continuous action, $\epsilon$ is added noise for exploration, and 
$d_\pi(s) $ is state distribution. 

To implement the TD3 algorithm, we need to define the reward for learning.
Using \eqref{opt2}, we define the reward as follows:
\begin{align}
\label{reward2}
r_t = -\ln\left( P_\text{out}\right).
\end{align}

The state is position of the UAV in 3D space denoted by $s(t) = [x_u(t),y_u(t),z_u(t)]$, which is a 3D continuous variable. Also, using \eqref{aa1}, action is a 3D continuous variable as
\begin{align} \label{ab1}
\left\{ \!\!\!\!\! \! \!
\begin{array}{rl}
	& a_x(t) = v_x(t)\Delta t + \frac{1}{2} v'_x(t) \Delta t^2, \\
	& a_y(t) = v_y(t)\Delta t + \frac{1}{2} v'_y(t) \Delta t^2, \\
	& a_z(t) = v_z(t)\Delta t + \frac{1}{2} v'_z(t) \Delta t^2. 
\end{array} \right.
\end{align}

Now, using the defined state, action and reward, we propose the Algorithm 1 to solve the optimization problem of the UAV trajectory.
In Algorithm 1, the variables $s''_0$ and $s'(t')$  indicate the initial and final position of the UAV in each trajectory. The UAV flies from point $s''_0$ to point $s'(t')$ based on the trajectory obtained from the Algorithm 1. The UAV stops at point $s'(t')$ until the network topology changes.

\begin{algorithm}
\caption{TD3-based trajectory algorithm}
\begin{algorithmic}[1]
	\renewcommand{\algorithmicrequire}{\textbf{Input:}}
	\renewcommand{\algorithmicensure}{\textbf{Output:}}
	\ENSURE  Trajectory $\mathcal{B}_u(t)$
	\\ \textit{Initialize all, critic networks $Q(s_t,a_t,\psi_1), Q(s_t,a_t,\psi_2)$ with $\psi_1,\psi_2$,
		actor network $\pi(s_t,\phi)$ whit $\phi$}
	\\ \textit{Initialize target networks $\psi'_1 \gets \psi_1$, $\psi'_2 \gets \psi_2$, $\phi' \gets \phi$ }
	\STATE {\it Initialize environment and reset $S_i$ for $i=\{1,...,M_s\}$.}
	\STATE{Generate random $T\in\{  T_\text{min},T_\text{max}\}$.}
	\FOR {episode = 1 to max-number-episodes}
	\STATE Observe the initial state $s(t)$
	\FOR {$n=1$ to max-episode-steps}
	\STATE Perform action $a(n) = \pi(s(n),\phi) + \epsilon$.
	\STATE Observe reward $r(n)$ and the next state $s(n+1)$.
	\STATE Store the transition ($s(n), a(n), r(n), s(n+1)$) in replay buffer.
	\STATE Sample mini-batch from replay buffer. 
	\STATE Update $\psi_1$ and $\psi_2$.
	\IF {$d_\text{del}$ mod $n$} 
	\STATE Update actor $\phi$.
	\STATE Update target networks: $\psi'_1 \gets \tau \psi_1 + (1-\tau)\psi'_1$, 
	$\psi'_2 \gets \tau \psi_2 + (1-\tau)\psi'_2$,
	$\phi' \gets \tau \phi + (1-\tau)\phi'$.
	\ENDIF
	\ENDFOR
	\ENDFOR
	\STATE Trajectory: Fly the UAV towards the end point. 
	\WHILE {$t<T$}
	\STATE Stay UAV at end point.
	\ENDWHILE
	\RETURN to line 1 
\end{algorithmic} 
\end{algorithm}

\section{Simulation Results}
For simulations, we consider that the UAV covers an area of $150 \times 150$ m$^2$. 
The UAV equipped with $M_u$ antennas at frequency $f c=140$ GHz.
One of the practical problems of using THz frequencies on UAVs is the power amplifier whose dimensions is large \cite{sarieddeen2021overview}. Therefore, we considered the maximum transmitted power of 10 mW for each antenna, which is practically possible for installation on a UAV.
To compensate the low transmitted power, we have used the array antennas on the UAV.
Each square array antenna includes of $N_{ui}\times N_{ui}=20\times 20$ elements with equal spacing $d=\lambda/2$ between elements. Therefore, the effective size of each array antenna is $A_\text{eff}\simeq\frac{18\times c}{f_c}=4.3$ cm \cite{balanis2016antenna}, which has suitable low dimensions for installation on the UAV. 
We have also assumed that the SBSs are distributed with a uniform random distribution and their topology changes every $T$ second. Parameter $T\in\{T_\text{min},T_\text{max}\}$ is also a random variable, where $T_\text{min}=20$, and $T_\text{max}=35$.
The maximum speed of the UAV is $v_\text{max}=8$ m/s, and its acceleration is $v'=4$ m/s$^2$. The UAV has a flight height limit of $h_\text{min}=30$ m, and $h_\text{max}=130$ m. Also, the intensity of the UAV's vibrations is considered $\sigma_\theta=2^o$.

For the machine learning configurations, all neural networks are initialized with the same parameters: each has two fully-connected hidden layers with size $256\times128$ neurons. 
Adam is used as the optimizer of both critic and actor networks. The hyper-parameters are set as follows: the learning rate of both the actor and critic networks are $10^{-4}$, the discount factor $\gamma= 0.9$, the mini-batch size 32, replay buffer size 1000, maximum number of episodes 150, the maximum steps per episode is 10 and the Poylal averaging factor $\tau=0.01$.

\begin{figure}
\begin{center}
	\includegraphics[width=3.2 in]{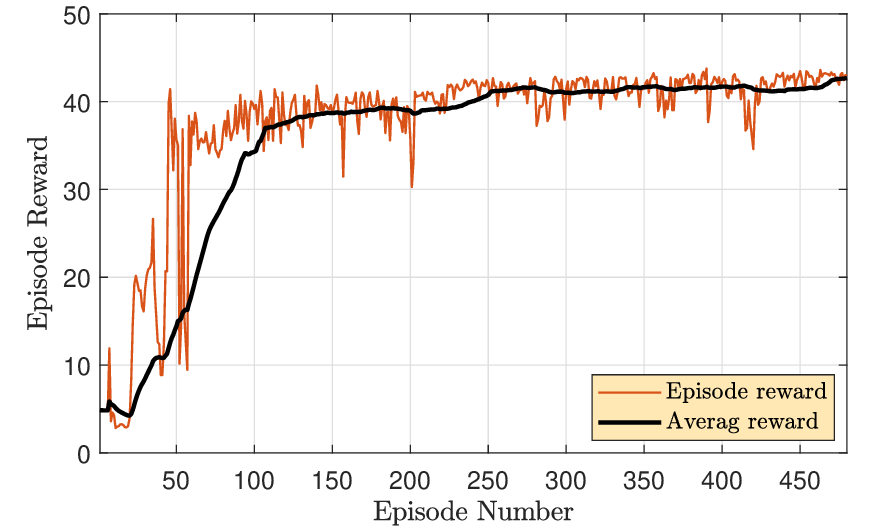}
	\caption{Episode and average reward versus episode number.}
	\label{he1}
\end{center}
\end{figure}

In order to show the convergence speed of the proposed algorithm, in Fig. \ref{he1}, we provide the obtained rewards after each episode in relation to the number of episodes during the learning process. In addition, the average rewards has been plotted by taking the average of the rewards obtained from 40 consecutive episodes. Based on the obtained results, it can be seen that the proposed algorithm almost converges after about 60 to 70 episodes. It should be noted that each episode contains 10 time steps. In total converges after about 600 to 700 time steps which is a fast and acceptable time for convergence.

%

\begin{figure}
\begin{center}
	\includegraphics[width=3.2 in]{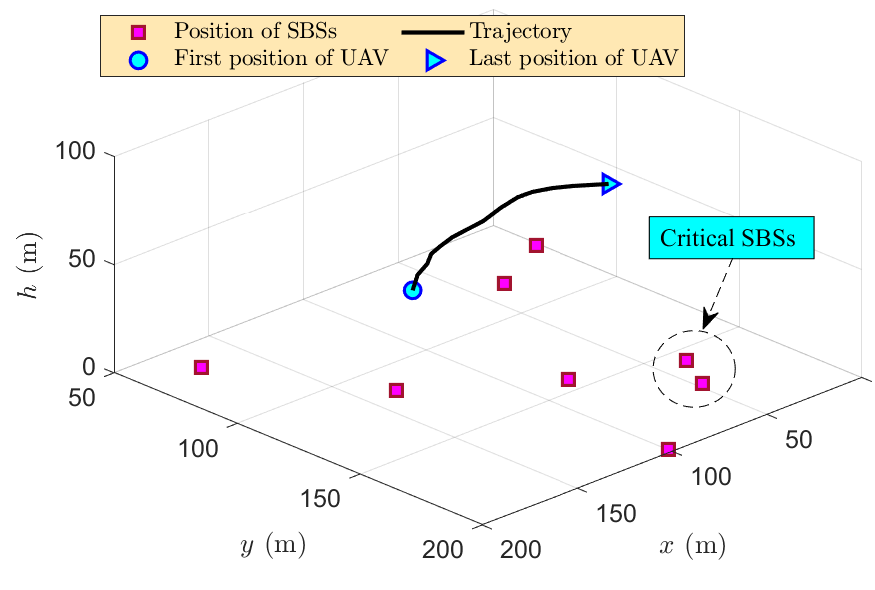}
	\caption{3D representation of the UAV trajectory for a random distribution of SBSs. The UAV found the critical SBSs and flies towards them to increase the spatial angle between the SBSs and decreases the interference.
	}
	\label{he2}
\end{center}
\end{figure}
%

\begin{figure}
\begin{center}
	\includegraphics[width=3.2 in]{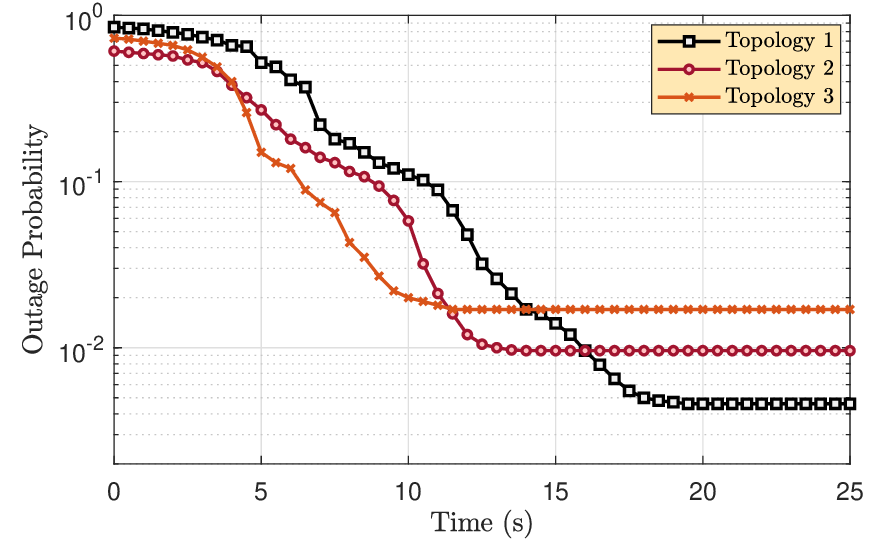}
	\caption{Variations of outage probability during UAV trajectory time for three different random distributions of SBSs.
	}
	\label{he3}
\end{center}
\end{figure}
%

To get a better understand, in Fig. \ref{he2}, the UAV trajectory obtained from the algorithm is plotted for a random distribution of SBSs. We defined the reward based on the minimize the maximum of outage probability.
Therefore, the UAV seeks to reduce the outage probability of the links with the highest outage probability. The highest outage probability is for the SBSs that are close to each other and cause the most interference. In this case, as we can see in Fig. \ref{he2}, the UAV flies towards two nodes close to each other (critical nodes) in such a way that the spatial angle between the nodes is maximized. To find more information about the importance of the trajectory, in Fig. \ref{he3}, the obtained outage probability (the minimum of the outage probability) along the trajectory is drawn for three different random distributions of SBSs. As it can be seen, with any random change in the network topology, the UAV quickly identifies the critical nodes and corrects its position in such a way that the outage probability is minimized.




\end{document}